\documentstyle[11pt,newpaspf,twoside,epsf]{article}
\def\edcomment#1{\iffalse\marginpar{\raggedright\sl#1\/}\else\relax\fi}
\reversemarginpar

\begin{document}
\title{Frequently Asked Questions on the Luminosity--Temperature
 relation in Groups and Clusters}
 \author{Fabio Governato}
\affil{Osservatorio Astronomico di Brera, Milan, Italy \\
 Astronomy Dept., University of Washington, Seattle, US}

\begin{abstract}
Efforts to understand the deviation of the L--T relation from a simple
scaling law valid for clusters and groups have triggered a number of
interesting studies on the subject. Techniques and approaches differ
widely but most works agree on the important role played by gas
cooling and heating sources like AGNs and SNe. Observations set useful
constraints on the evolution of the intracluster medium (ICM): a
100KeV/cm$^2$ entropy floor in the core of groups and about 5--15\% of
baryons being converted into stars. However, essential details like
the nature of the dominant heating mechanism and the quantitative
importance of cooling still need to be addressed. I suggest that a new
generation of high resolution N-body simulations and a quantitative
comparison of results between different approaches is required to
improve results and increase our understanding of the problem.

\end{abstract}

\section{Q: What are the Theoretical Expectations for the L--T relation?}

The first attempt to model the ICM in the framework of the
hierarchical scenario assumed its thermodynamical properties to be
entirely determined by gravitational processes, like adiabatic
compression during collapse and shock heating (Kaiser 1986). If no
characteristic scales are present in the underlying cosmology (i.e.,
Einstein--de-Sitter cosmology and power--law shape for the power
spectrum of density perturbations), this model should predict hot gas
within rich clusters to look the same as within poor groups, since
gravity in itself does not have characteristic scales. Under the
assumptions of emissivity dominated by free--free bremsstrahlung and
of hydrostatic equilibrium of the gas, this model predicts $L_X\propto
T^2(1+z)^{3/2}$ for the shape and evolution of the relation between
$X$--ray luminosity and ICM temperature (also Eke, Navarro \& Frenk
1998). Furthermore, if we define the gas entropy as $S=T/n_e^{2/3}$
($n_e$: electron number density; e.g., Eke et al. 1998), then the
self--similar ICM has $S\propto T(1+z)$.

\section{Q: What are the observational Constraints?}

However, soon the simple model described above failed to account for
several new observational facts: (a) the $L_X$--$T$ relation for
nearby clusters is steeper than predicted, with $L_X\propto T^{\sim
3}$ for $T> 2$ keV clusters (e.g., David et al. 1993; White, Jones \&
Forman 1997; Allen \& Fabian 1998; Markevitch 1998); with a possible
further steepening at the group scale, $T < 1$ keV (Ponman et
al. 1996; Helsdon \& Ponman 2000); (b) no evidence of evolution for
its amplitude has been detected out to $z> 1$ (e.g., Fairley et
al. 2000; Della Ceca et al. 2001; Borgani et al. 2001a); (c) the gas
density profiles in central regions of cooler groups is relatively
softer than for hotter cluster and, correspondingly, the entropy is
higher (e.g., Ponman, Cannon \& Navarro 1999; Lloyd--Davis et
al. 2000).  Observational evidence provides a fairly robust measure of
the entropy floor for the diffuse hot gas in groups (Lloyd--Davis et
al. 2000) of 100 KeV/cm$^2$.

\noindent A related, very important measurement is the fraction of
baryons that has been converted into a ``cold'', non X--ray emitting
phase (i.e stars and neutral hydrogen). This constraint is crucial as
it defines the amount of energy available from SNe and the importance
of cooling in groups and clusters (Renzini 2000). Balogh et al. (2001)
estimate that only a small fraction (5--10\%) of baryons has been
converted into stars, almost independently of the total virial mass of
the cluster.

\section{Q: What are the physical processes that could affect the L--T relation?}

$\bullet$ {\it Gas cooling}. At the center of groups the cooling time
is much shorter than a Hubble time. Cooling would remove low entropy
gas transforming it into stars and possibly originate the entropy
floor observed.  While advocated on the ground of its simplicity
(Muanwong et al 2001 also Voit \& Bryan 2001), the amount of gas
involved cannot exceed the observed fraction of 5--15\% of the total
baryon fraction.

\noindent  
$\bullet$ {\it Energy injection from AGNs} (e.g B\"ohringer, this
proceedings) would be an important source of heating. Most of the
energy would come from jets that would carve ``bubbles'' in the ICM
(Mc Namara et al 2001, Quilis et al. 2001), while the X-ray emission
from the central engine in radio quiet QSOs would likely have a small
effect due to the small cross section for the process.

\noindent
$\bullet$ {\it Energy injection from SNe}. Winds propelled by SNe
explosions (an hefty 10$^{51}$erg/SN) would transfer heat to the ICM,
possibly escaping the halos of individual galaxies in the forming
protocluster and enrich the ICM with metals.  (Note: substantial metal
enrichment could originate also from ram pressure and tidal stripping of
galaxies (Moore, Quilis \& Bower 2001).

\section{Q: What are the most used theoretical tools to study the L--T relation?}

\noindent
$\bullet$ {\it Analytical methods} allow a fast exploration of
parameter space, but must sometimes rely on the parametrization of the
end results of complicated physical processes. While this approach
requires studying the physics of gas in a somewhat idealized situation
(spherical collapse, accretion from a uniform background) the amount
of insight gained is impressive, although conclusions sometimes differ
widely (Kaiser 1991,Valageas \& Silk 1999, Tozzi \& Norman 2000, Bryan
2000, Balogh, Babul \& Patton 1999, among many). The most refined
methods are based on  the standard machinery developed to link
gravitational clustering in Cold Dark Matter Models and the physics of
gas and star formation (Cole et al. 2000, Somerville \& Primack 1999,
Bower et al. 2001, Menci \& Cavaliere 2000, Cavaliere, Giacconi \&
Menci 2000).

\noindent
$\bullet${\it N-body simulations} are a powerful method, able to treat
the highly non--linear problem of the formation of cosmic structures
with very few assumptions. Recent observational evidence for a mono
phase ICM (B\"ohringer et al 2001) remove some potential worries about
the use of SPH methods, which do not describe well a multiphase
medium. Some recent works on the L--T relation including feedback and
star formation (Metzler \& Evrard 1994, Bialek, Evrard \& Mohr 2001,
Borgani et al 2001b, Valdarnini 2001) provide a good starting point
and an extensive list of references to the subject of simulations of
galaxy clusters.

\section{Q: Heating, Cooling or both?}

As mentioned above, several papers suggested that simply the addition
of cooling would remove enough gas from the ICM to reconcile models with the
observed L--T relation. This would likely be in contrast with
observational evidence that points to a limited fraction of baryons in
clusters in a ``cold'' non X--ray emitting phase.  Moreover it would
be difficult to hide the large quantity of cooled gas in ``dark
baryons'' as they would cause detectable deepening of the central
potential well at the center of clusters (e.g Lewis et al. 1999,
Valdarnini 2001). This would have a substantial effect on the
predicted M--T relation increasing the emission weighted temperature
of groups (Finoguenov et al 2001). On the other hand recent
theoretical works, both analytical and numerical seem to (perhaps
slowly) converge to a required energy input comparable to 100\% of
what is available from SNe (Kravtsov \& Yepes 2000, Bower et al 2001, or
even a few times higher (Borgani et al. 2001b, Valageas \& Silk 1999,
Wu, Fabian \& Nulsen 2000) to satisfy the L--T relation.  While
semi-analytical models estimate the energy available from SNe to be of
the order of $\sim$ 0.25 keV per particle , Renzini (2000) suggested
an even smaller energy budget available from SNe: just 0.1 keV. Taken
together these works suggest that the required energy must be provided
by both SNe and AGNs, perhaps active as major mergers shape the early
type population of cluster galaxies.  Unfortunately none of the
N--body simulations cited above (some analytical works did) treated
{\it both} cooling and heating in a fully consistent way or included
the effect of AGNs on the ICM. This leaves the quantitative
contribution of cooling and heating rather uncertain.

\section{Q: Is there a preferred epoch for heating the ICM?}

Not really. As the entropy S depends on a inverse power
 of $\rho$ it might seem more efficient to heat the ICM at lower z
 (say z$<5$) as suggested in Tozzi \& Norman (1999). However, in their
 model the authors heated the gas {\it before} accretion, while at the
 background density.  In a more realistic case gas will be heated when
 already inside halos that will later merge to form the cluster. While
 this requires more energy for a given entropy S, ($\rho$ is higher by
 a factor $\sim 200$) this is more than compensated by the fact that
 halos collapsed at high z will form the core of the cluster at lower
 z (Governato et al. 2001), i.e. you dump energy right where it is
 needed.  Borgani et al (2001b, 2001c) and Bialek, Evrard \& Mohr
 (2000) tried a large range of z (1 to 5 and 20 respectively) finding
 a weak dependence on results for the argument given above.  Of
 course, a redshift too close to z $\sim $1 for heating would cause
 the L--T to evolve significantly at moderate redshift, contrary to
 observations.

\noindent
Bottom line: it is very likely that a combination of
cooling \& feedback will give the correct, non--evolving L--T relation,
although only stronger observational constraints and more refined
simulations will allow us to evaluate the {\it quantitative} role of
cooling, SNe and AGNs.

\begin{figure}  
\plotone{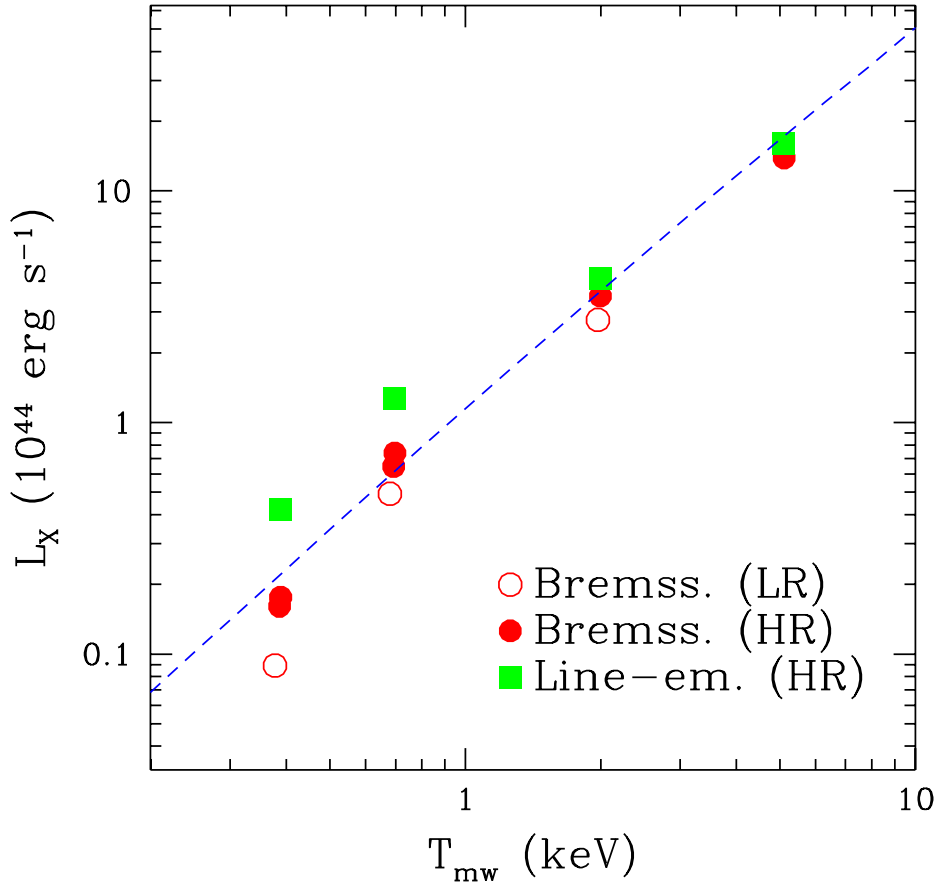}
\caption{X-ray luminosity vs mass weighted Temperature: effects of
resolution.  Empty (red) dots are runs with $\sim$ 0.5--2 $\times$
10$^4$ gas particles within R$_{vir}$. Full dots are halos simulated
with N$_{gas}$$(r>$R$_{vir}$) $\sim$ 10$^5$. Softening is
1$\%$R$_{vir}$.  The upper filled (red) dots show the effect of having
instead $\epsilon$ = 0.5$\%$ R$_{vir}$: L$_x$ grows by 15--20\%.
(Green) filled squares show the large effect of including line
emission in the estimate of X--ray luminosities for the high res
runs. The dashed line is the scaling predicted by Kaiser (1986)
assuming NFW (Navarro, Frenk \& White 1995) profiles. ( data from
Borgani et al. 2001 in prep.){\it Bottom line: use at least 10$^5$ gas
particles per halo to avoid nasty systematic effects like reproducing
a bend in the L--T relation at T $<$ 0.3 KeV. Correct for line
emission if you compare vs. real data.}  
}
\end{figure}

\section{Future Improvements:}

Numerical simulations including a treatment of hydrodynamical
processes have reached a high degree of maturity, with widely
different codes giving comparable results (Frenk et al 1999). However,
the number of particles used to simulate {\it individual} halos is
still systematically lower than in DM only runs (e.g Ghigna et
al. 2001).  The hot debate on the shape of inner density profiles in
DM halos has started a race to higher resolution runs (Governato,
Ghigna \& Moore 2001). While the quest for the ultimate density
profile of DM halos might ultimately prove futile it has sparked a
number of improvements on codes and, equally important, increased the
robustness of results.  A key method is the so called
``renormalization technique'' (Katz \& White 1993) where a halo taken
from a cosmological simulation is resimulated at much higher
resolution, while sparsely resampling the surrounding region.  Fig.1
shows the crucial effect of increasing resolution on a sample of SPH
simulations of renormalized halos. (Borgani et al 2001b,c). The L--T relation
obtained from simulations of large cosmological volumes (where
particle number and the ratio softening/R$_{vir}$ for each halo depends on its
virial mass) likely suffer of systematic biases due to the worsening
resolution for less massive halos: namely smaller L$_x$ and slower
cooling due to lower central densities.

\vskip 0.3truecm
\noindent
This is my very personal wish list for future theoretical and
observational studies on the subject:
\vskip 0.2truecm

\noindent
$\bullet$ Estimates of L$_X$ and T  in  groups that exclude
 contamination from AGNs.

\noindent 
$\bullet$ Estimates of the amount of metals in clusters to a
significant fraction of R$_{vir}$ as another method to infer the
energy release from SNe (Renzini 2000, De Grandi \& Molendi 2001).

\noindent 
$\bullet$ A robust estimate of the fraction of ``cold'' baryons which
  keeps into account the amount of intracluster stars. Recent claims
  (see Arnaboldi et al 2001) based on intracluster PNs  almost
  double the the efficiency of star formation in clusters.

\noindent
$\bullet$ Simulations of sufficient resolution to treat cooling and
star formation processes in a robust way. (note: robust does not mean
realistic, but it's a start).

\noindent
My checklist for simulating an individual cluster:

\vskip 0.5truecm

\noindent
  N$>$10$^5$ \\
  $\epsilon$/R$_{vir} < 1\%$ \\ 
  N$_{gas}(r<$R$_{vir})>$ 10$^5$ \\
  Simulation box $\sim$ 100Mpc \\
  N$_{steps}$ $>>$ 10$^{3-4}$ \\

\noindent
$\bullet$ Exploring the role of QSO's and AGNs in heating the
  ICM in a cosmological context.

\noindent
$\bullet$ Introducing heating schemes that closely  follow
  the time evolution of the physical process behind it.

\noindent
$\bullet$ Resolving the discrepancy between adiabatic runs with mesh
and SPH codes (e.g Kravtsov et al. 2001, also Springel \& Hernquist
2001 and Cen \& Ostriker 1999) : why SPH simulations do not have an
entropy floor and mesh codes do?

\acknowledgments

I wish to thank Stefano Borgani, Marino Mezzetti and Riccardo
Valdarnini for organizing such an interesting and lively meeting and
Uros Seljak and Bepi Tormen for stimulating discussions (although they
made me skip a coffee break or two).  I also thank my collaborators
Stefano Borgani, Nicola Menci, Tom Quinn, Joachim Stadel, Paolo Tozzi
\& James Wadsley for allowing me to discuss our results and show
Fig.1. in advance of publication.

\section{References}

Allen S.W., Fabian A.C., 1998, \mnras, 297, 63 \\
Arnaboldi et al, 2001 AJ in press \\
Balogh M.L., Babul A., Patton D.R., 1999, \mnras, 307, 463 \\
Balogh et al., 2001, \mnras, 326, 1228 \\
Bialek, J.J., Evrard, A.E., Mohr, J.J, 2001, \apj, 555, 597 \\
B\"ohringer, H. et al 2001, A\&A, in press \\
Borgani et al. 2001a \apj, 561, 13 \\
Borgani et al. 2001b, \apjl, 559, 71 \\ 
Borgani et al. 2001c, in preparation \\
Bower R.G., Benson A.J., Bough C.L., Cole S., Frenk C.S., 2001 \mnras
 ~325, 497 \\
Bryan G.L. 2000, \apjl, 544, 1 \\
Cavaliere A., Giacconi R., Menci N., 2000, \apj, 528, L77 \\
Cen R., \& Ostriker J.P.,  1999 \apj 517, 31 \\
Cole, S., Lacey, C.~G., Baugh, C.~M., \& Frenk, C.~S.\ 2000, \mnras, 319, 168\\
David L.P., Slyz A., Jones C., Forman W., Vrtilek S.D., Arnaud K.A., 1993, \apj, 412, 479  \\
De Grandi S. \& Molendi S., 2001 \apj, 551, 153 \\ 
Della Ceca R. et al. 2000, A\&A, 353, 498  \\
Eke, V., Navarro J.F. \& Frenk C.S. 199,8 \mnras, 503, 569 \\
Fairley  B.W. et al., 2000, \mnras, 315, 669 \\
Finoguenov A., Reiprich T.H., B\"ohringer H., 2001, A\&A \\
Frenk, C.S. et al. 1999, \apj,  525, 554 \\
Ghigna S. et al. 2000, \apj, 544, 616 \\
Governato, F., Ghigna, S. \& Moore, B. 2001, proceedings of
``Astrophysical Ages and Timescales, ed. Ted Von Hippel, (astro-ph/0105443)\\
Governato, F., Ghigna, S., Moore, B., Quinn, T., Stadel, J., \& Lake,
G.\ 2001, \apj, 547, 555 \\
Helsdon S.F., Ponman T.J., 2000, \mnras, 315, 356 \\ 
Kaiser N., 1986, \mnras, 222, 323 \\
Kaiser N., 1991, \apj, 383, 104 \\
Katz, N. \& White, S.D.M., 1993 \apj412, 455 \\
Kravtsov A.V., Yepes G., 2000, \mnras, 318, 227 \\
Kravtsov A.V. et al. \apj~~ submitted, (astro--ph/0109077) \\
Lewis G.F., Babul A., Katz N. Quinn T., Hernquist L., 2000, \apjl, 536, 623 \\
Lloyd-Davies E.J., Ponman T.J., Cannon D.B., 2000, \mnras, 315, 689 \\
Markevitch M., 1998, ApJ, 504, 27 \\
Mc Namara, B.R. et al 2001, \apjl, 562, 149 \\
Muanwong et al 2001, \apjl 2001 552, 27 \\
Menci N., Cavaliere A., 2000, \mnras, 311, 50 \\
Metzler C.A., Evrard, A.E. 1994 \apj 437, 564  \\ 
Moore B., Quilis V. \& Bower, R. 2000, ASP  Conf. Sr. 197, 363 \\
Ponman T.J., Bourner P.D.J., Ebeling H., B\"ohringer H., 1996, 293, 690 \\
Ponman T.J., Cannon D.B., Navarro J.F., 1999, Nature, 397, 135 \\
Quilis, V., Bower, R.G. \& Balogh, M.L. 2001,  astro-ph/0109022 \\
Renzini A., 2000 astro-ph/0001312, proccedings of ``LSS in the X--ray Universe, 1999, ed Atlantisscience. \\
Somerville, R.S., \& Primack, J.R. 1999, \mnras, 310, 1087 \\
Springel, V. \& Hernquist L. 2001 astro-ph/0111016 \\
Tozzi P., Norman C. 2001, \apj, 546, 63 \\
Valageas P., Silk J., 1999, A\&A, 350, 725 \\
Valdarnini R., 2001, \apj, in press astro--ph/0110545  \\
Voit G.M.,  \& Bryan G.L., 2001, Nature, in press \\ 
White D.A., Jones C., Forman W., 1997, \mnras, 292, 419\\
Wu K.K.S., Fabian A.C., Nulsen P.E.J., 2000, \mnras, 318, 889 \\

\end{document}